\documentclass{article}
\usepackage{amsmath}
\usepackage{chicago}

\input{tcilatex}
\begin{document}

\title{$\mathbf{Phase}$ $\mathbf{structure}$ $\mathbf{of}$ $\mathbf{%
Topologically}$ $\mathbf{massive}$ $\mathbf{gauge}$ $\mathbf{theory}$ $%
\mathbf{with}$ $\mathbf{fermion}$}
\author{Yuichi Hoshino \\
Kushiro National College of Technology,\\
Otanoshike-nishi 2-32-1,Kushro 084-0916,Hokkaido,Japan}
\maketitle

\begin{abstract}
Using Bloch-Nordsieck approximation fermion propagator in 3-dimensional
gauge theory with topological mass is studied.Infrared divergence of
Chern-Simon term is soft,which modifes anomalous dimension.In unquenched QCD
with 2-component spinor anomalous dimension has fractional value,where order
parmeter is divergent.
\end{abstract}

\section{ Introduction}

The Lagrangeans of Topologically massive gauge theory with fermion are\cite%
{jackiw}%
\begin{equation}
L=\frac{1}{4}F_{\mu \nu }F^{\mu \nu }+\frac{1}{4}\theta \epsilon ^{\mu \nu
\rho }F_{\mu \nu }A_{\rho }+\overline{\psi }(i\gamma \cdot (\partial
-ieA)-m)\psi +\frac{1}{2d}(\partial \cdot A)^{2},
\end{equation}%
\begin{align}
L& =\frac{1}{4g^{2}}tr(F_{\mu \nu }F^{\mu \nu })-\frac{\theta }{4g^{2}}%
\epsilon ^{\mu \nu \rho }tr(F_{\mu \nu }A_{\rho }-\frac{2}{3}A_{\mu }A_{\nu
}A\rho )  \notag \\
& +\overline{\psi }(i\gamma \cdot (\partial -ieA)-m)\psi +\frac{1}{2d}%
(\partial \cdot A)^{2}
\end{align}%
,where $\theta =(g^{2}/4\pi )n,(n=0,\pm 1,\pm 2...)$ with $\gamma $ matrix
for 4-comonent fermion \cite{Kondo}.In comparison with massless $%
QED_{3},QCD_{3}$ topological mass term seems to soften the infrared
divergence of massive fermion near its on-shell.In Minkowski metric we have
5 $\gamma $ matrices $\{\gamma _{\mu },\gamma _{\nu }\}=2g_{\mu \nu },(\mu
=0,1,2)$%
\begin{align}
\gamma _{0}& =\left( 
\begin{array}{cc}
\sigma _{3} & 0 \\ 
0 & -\sigma _{3}%
\end{array}%
\right) ,\gamma _{1,2}=-i\left( 
\begin{array}{cc}
\sigma _{1,2} & 0 \\ 
0 & -\sigma _{1,2}%
\end{array}%
\right) ,\gamma _{4}=\left( 
\begin{array}{cc}
0 & I \\ 
I & 0%
\end{array}%
\right) ,\gamma _{5}=\left( 
\begin{array}{cc}
0 & -iI \\ 
iI & 0%
\end{array}%
\right) ,  \notag \\
\tau & \equiv \frac{-i}{2}[\gamma _{4},\gamma _{5}]=diag\left( I,-I\right)
,\tau _{\pm }=\frac{1\pm \tau }{2}.
\end{align}%
There are two redundant matrices $\gamma _{4\text{ }}$and $\gamma _{5}$
which anticommutes with other three $\gamma $ matrices$.$There exists two
kind of chiral transformation $\psi \rightarrow \exp (i\alpha \gamma
_{4})\psi ,\psi \rightarrow \exp (\alpha \gamma _{5})\psi .$The matrices $%
\{\gamma _{4},\gamma _{5},I_{4},\tau \}$ generate a $U(2)$ chiral symmetry
containing massless spinor fields.This $U(2)$ symmetry is broken down to $%
U(1)\times U(1)$ by a spinor mass term $m_{e}\overline{\psi }\psi $ and
parity violating mass $m_{o}\overline{\psi }\tau \psi ,$where $\overline{%
\psi }\tau \psi $ is a spin density.Here after we take 4-component spinors
to study chiral symmetry breaking by $m^{e}\overline{\psi }\psi $ in pure QED%
$_{3},$where $\overline{\psi }\tau \psi $ is invariant under chiral
transformation.After that the effects of Chern-Simon term will be studied by
2-component spinors.

\section{Fermion spectral function}

The spectral function of 4-component fermion and photon propagator\cite{Solo}
are defined as 
\begin{equation}
S_{F}(x)=S_{F}^{0}(x)\exp(F(x)),S_{F}^{0}(x)=-(i\gamma\cdot\partial +m)\frac{%
\exp(-m\sqrt{-x^{2}})}{4\pi\sqrt{x^{2}}}.
\end{equation}%
\begin{equation}
D_{F}^{0}(k)=-i(\frac{g_{\mu\nu}-k_{\mu}k_{\nu}/k^{2}-i\theta\epsilon_{\mu
\nu\rho}k^{\rho}/k^{2}}{k^{2}-\theta^{2}+i\epsilon})+id\frac{k_{\mu}k_{\nu}}{%
k^{4}},
\end{equation}
where $F$ is an $O(e^{2})$ matrix element $|T_{1}|^{2}$ for the process
electron $(p+k)\rightarrow$electron $(p)+$photon $(k)$ as 
\begin{align}
T_{1} & =-ie\frac{\epsilon_{\mu}(k,\lambda)}{\gamma\cdot(p+k)-m}\gamma^{\mu
}\exp(i(p+k)\cdot x)U(p,s), \\
\sum_{\lambda,S}T_{1}\overline{T_{1}} & =-\frac{\gamma\cdot p+m}{2m}e^{2}[%
\frac{m^{2}}{(p\cdot k)^{2}}+\frac{1}{p\cdot k}+\frac{(d-1)}{k^{2}}]-\frac{%
\gamma\cdot p}{m}\frac{e^{2}}{4\theta}\frac{m}{p\cdot k},
\end{align}%
\begin{equation}
F=\int\frac{d^{3}k}{(2\pi)^{2}}\exp(ikx)\theta(k_{0})\delta(k^{2}-\theta
^{2})\sum_{\lambda,S}T_{1}\overline{T_{1}}.
\end{equation}
Here we use the retarded propagator to derive the function $F$ 
\begin{equation}
D_{+}(x)=\int\frac{d^{3}k}{i(2\pi)^{2}}\exp(ik\cdot
x)\theta(k_{0})\delta(k^{2}-\theta^{2})=\frac{\exp(-\theta\sqrt{-x^{2}})}{%
8\pi i\sqrt {-x^{2}}}
\end{equation}
The function $F$ is evaluated by $\alpha$ integration for pure QED$_{3}$\cite%
{YH,Solo}. 
\begin{align}
F & =ie^{2}m^{2}\int_{0}^{\infty}\alpha d\alpha D_{F}(x+\alpha
p)-e^{2}\int_{0}^{\infty}d\alpha D_{F}(x+\alpha p)-i(d-1)e^{2}\frac{\partial%
}{\partial\theta^{2}}D_{F}(x,\theta^{2})  \notag \\
& =\frac{e^{2}}{8\pi}[\frac{\exp(-\theta|x|)-\theta|x|E_{1}(\theta |x|)}{%
\theta}-\frac{E_{1}(\theta|x|)}{m}+\frac{(d-1)\exp(-\theta\left\vert
x\right\vert ))}{2\theta}].
\end{align}
It is well known that function $F$ has linear and logarithmic infrared
divergence with respect to $\theta$ where $\theta$ is a bare photon mass$.$%
Here we notice the followings.(\textbf{1) }$\exp(F)$ includes all infrared
divergences$.$(\textbf{2)} quenhed propagator has linear and logarithmic
infrared divergences.Linear divegence is absent in a special gauge.%
{\normalsize In unquenched case }$\theta$ dependence of $\exp(F)$ is
modified {\normalsize by} dressed boson spectral function to $\int
ds\rho_{\gamma}(s)\exp(F(x,\sqrt{s}),$where $\pi\rho_{\gamma}(s)=-\Im
(s-\Pi(s))^{-1}.$

\section{Phase strucutures}

For short and long distance we have the approximate form of the function $F$ 
\begin{equation}
F_{S}\sim A-\theta|x|+(D+C|x|)\ln(\theta\left\vert x\right\vert ))-\frac {%
(d+1-2\gamma)e^{2}|x|}{16\pi},(\theta|x|\ll1),F_{L}\sim0,(1\ll\theta|x|)..
\end{equation}
From the above formulae we have 
\begin{align}
\exp(F) & =A(\theta|x|)^{D+C|x|}(\theta|x|\ll1),A=\exp(\frac{e^{2}(1+d)}{%
16\pi\theta}+\frac{e^{2}\gamma}{8\pi m}),  \notag \\
C & =\frac{e^{2}}{8\pi},D=\frac{e^{2}}{8\pi m},
\end{align}
where $m$ is the physical mass and $\gamma$ is an Euler's constant.Here we
see $D$ acts to change the power of $|x|$.For $D=1,S_{F}(0)$ is finite and
we have $\left\langle \overline{\psi}\psi\right\rangle \neq0$ \cite{YH,maris}%
.Thus if we require $D=1,$we obtain $m=e^{2}/8\pi.$In the same way we add
the conrtribution of Chern-Simon term.For simplicity we consider the
2-component spinors in(7).In the condensed phase we have the modified
anomalous dimension for $\theta>0$ 
\begin{equation}
D=\frac{e^{2}}{8\pi m}+\frac{e^{2}}{32\pi\theta}=1,m^{^{\prime}}=\frac{e^{2}%
}{8\pi}/(1-\frac{e^{2}}{32\pi\theta}),\Delta m=\frac{e^{2}}{8\pi}\frac{e^{2}%
}{32\pi\theta}/(1-\frac{e^{2}}{32\pi\theta}).
\end{equation}
In unquenched case there are parity even and odd spectral function of gauge
boson by vacuum polarization\cite{jackiw}.In this case we separate $D$ into
parity even and odd contribution $D^{e}=e^{2}/8\pi m,D^{O}=e^{2}/32\pi\theta
.$For Topologically Massive QCD,$\theta$ is quantized with $n(0,\pm
1,\pm2,..).$Thus we have $D=e^{2}/(8\pi m)+1/8n=1$ for quenched case, and $%
D^{e}=e^{2}/8\pi m=1,D^{O}=1/(8n)$ for unquenched case,which leads to $%
\left\langle \overline{\psi}\psi\right\rangle =\infty$ for any $n\neq0.$In
the 4-component spinor free fermion propagator is decomposed into chiral
representation

\begin{equation}
S_{F}(p)=\frac{1}{m_{\epsilon}I+m_{O}\tau-\gamma\cdot p}=\frac{(\gamma\cdot
p+m_{+})\tau_{+}}{p^{2}-m_{+}^{2}+i\epsilon}+\frac{(\gamma\cdot
p+m_{-})\tau_{-}}{p^{2}-m_{-}^{2}+i\epsilon}.
\end{equation}
The difference in two spectral functions is a opposite sign of each
Chern-Simon contribution for $\tau_{\pm}$.In Toplogically massive gauge
theory dynamical mass is parity even and Chern-Simon term shifts mass of
different chirality with opposite sign.However shifted mass may not strongly
depend on $\theta$ but $\left\langle \overline{\psi}\psi\right\rangle $ is
proportional to $\theta$(12)\cite{matsu}.Our approximation is convenient for
unquenched case by the use of gauge boson spectral function\cite{maris}.

\noindent

\end{document}